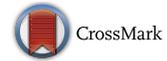

# Graph-Based Entropy for Detecting Explanatory Signs of Changes in Market

Yukio Ohsawa[1]

**Abstract**
Graph-based entropy, an index of the diversity of events in their distribution to parts of a co-occurrence graph, is proposed for detecting signs of structural changes in the data that are informative in explaining latent dynamics of consumers' behavior. For obtaining graph-based entropy, connected sub-graphs are first obtained from the graph of co-occurrences of items in the data. Then, the distribution of items occurring in events in the data to these sub-graphs is reflected on the value of graph-based entropy. For the data on the position of sale, a change in this value is regarded as a sign of the appearance, the separation, the disappearance, or the uniting of consumers' interests. These phenomena are regarded as the signs of dynamic changes in consumers' behavior that may be the effects of external events and information. Experiments show that graph-based entropy outperforms baseline methods that can be used for change detection, in explaining substantial changes and their signs in consumers' preference of items in supermarket stores.

**Keywords** Graph-based entropy · Explanatory signs of change · Marketing

## 1 Introduction

Statistics and machine learning have been adopted to forecasting demands in markets [1, 2]. However, changes in the market due to the effects of external events are hard to explain by learning causalities from data, because external causal events are out of data by definition. Here, we define explanation as to relate a change in the observation with causes that may not be events in the data. Let us assume that we have data on the position of sale (POS) in a supermarket as $D$ in Eq. (1), that is the target data dealt with in this paper, where $B_t$ stands for a basket, i.e., a set of items (members of $I$, the set of all items in the supermarket)

✉ Yukio Ohsawa
ohsawa@sys.t.u-tokyo.ac.jp

[1] Department of Systems Innovation, School of Engineering, The University of Tokyo, 7-3-1 Hongo, Bunkyo-ku, Tokyo 113-8653, Japan

            ≰ Springer

purchased at time $t$ by some consumer and $T$ is the length of the period of time in the data:

$$D = \{B_t | 0 < t \leq T\}. \tag{1}$$

Here, suppose that the sales' volume of coffee increases for a week beyond prediction on the POS data. The cause of this increase may be a TV program broadcasted a few days ago, about the positive effect of coffee on human's health. Such a causality of change may be explained if a marketer focuses attention on the period of time when the external cause, i.e., the TV program about healthcare that is not in set *I*, occurred and if additional data about past TV programs are given. Then, the marketer can create a strategy to promote the sales of coffee by publishing a book relevant to the content of the TV program.

Change points have been detected on the changes in parameters and/in models of time series in the approach of machine learning. In Principal Component Analysis (PCA), projecting data to principal components do not only reduce computational cost, but also sharpen the sensitivity of change detection. Here, the change in the correlation, the variance, and the mean of components is detected from before to after a change [3]. Methods for detecting changes in parameters in the model capturing the structure of latent causality have been developed for both discrete [4–6] and continuous [7] changes, and the method for the latter is turning out to work for the former as well. The changes in the values of the parameter set $\Theta$, from time $t - \delta t$ to $t$, are learned as $\Theta[t - \Delta t, t] - \Theta[t - \delta t - \Delta t, t - \delta t]$, where $\Delta t$ is the width of the training time window of the data to learn $\Theta[t - \Delta t, t]$ and $\Theta[t - \delta t - \Delta t, t - \delta t]$ from, and $\delta t$ is the time step of the change. $\Theta$ is learned to minimize the error of prediction from the reality of observable events. The precision of change detection is expected to be the better for the larger $\Delta t$ that can be regarded be a part of tolerant delay, i.e., the length of time the analysis should wait for detecting a change. However, a large $\Delta t$ is not reasonable from the viewpoint to explain the change quickly. For example, to highlight the causality above from the transient TV program to the coffee sales, $\Delta t$ is better to be set to 1 month than to 1 year, so that the TV program can outstand as an essential cause in the period of length $\Delta t$.

From to the viewpoint not only to detect, but also to explain a change with linking to external knowledge, i.e., knowledge about events not included in data, there are methods to learn latent topics of interest in a sequence of words or actions without known labels corresponding to the topics. For example, consecutive time segments, each of which is relevant to a vector in the space of a limited number of latent topics that are not labeled by known labels, are obtained by the dynamic topic model (DTM [8]). By applying DTM to POS data, the changes in consumers' interests can be detected as the boundaries between the obtained time segments corresponding to the changes in the topic vector. Topic-tracking model (TTM) has been also presented to consider the evolution of each consumer if the behavior of each consumer $c$ is reflected on $D$ in the form of $B_{t,c}$ instead of $B_t$ in Eq. (1) regarding each consumer as a generator of topic vectors [9]. Topic models have a potential not only to learn topics behind observed events, but also to explain changes. In contrast, the aim of this paper can be positioned as to cope with changes, where such a transient topic, as the healthy coffee above, causes influence on the market and may disappear or get united with other topics.

Furthermore, to explain a change as an effect of an external cause, it is essential to detect a precursor that may be an evidence of the causality. In the example above, the precursor of the increase in the sales of coffee may be a novel co-occurrence of coffee with some healthy food in consumers' purchase, because people interested in health care may be the leading users of coffee. Such a precursor should appear in a short period that is before a larger number of people start to buy coffee but is after the TV program. Thus, this paper is addressed to the problem to detect a sign of change, i.e., any evidence of the change or the precursor of the change, on the data of a short $\Delta t$ and also to explain the sign with linking to external events.

Precursors to changes have been really explored in various domains, such as epileptic seizure [10], natural phenomenon [11], aviation [12], etc. These studies aim at alerting to a predefined influential event early enough for prevision, treatment, or management of consequences. Because of the requirement to explain what is coming after the precursor and what human(s) should do, these approaches directly or indirectly use knowledge and models of the dynamics of events in the target domain. For example, the state transition in disease progress [10] has been modeled for detecting precursory symptoms, and physical models have been used for monitoring events relevant to future earthquakes [11]. Methods for precursor detection have been developed also in aviation, to enable human–machine interaction for managing anomalous events [12, 13]. The strength of thus linking external knowledge, out of the data in the target of analysis, is twofold: the potential to explain causalities and the reinforcement of sensitivity to precursors. The idea to use external information for detecting and explaining precursors is also found in extracting associative relations between drugs and symptoms from the text in online medical forums that appear before the changes in label drugs by the Food and Drugs Associations [14].

In this paper, we first set a rough model of consumers' preference transition in the market, which may be caused by external events, i.e., events out of data in Sect. 2. This transition occurs from seeking diversity to focusing on preferred items, and vice versa, during which new interests of consumers may emerge, possibly due to influential external events. In Sect. 3, we define graph-based entropy (GBE in short) as an index for detecting structural changes of events' occurrence, modeled as the changes of subgraphs that are graph-based clusters. Such structural changes are regarded as a computational model of the transitions of consumers' preference that is redefined as a context in Sect. 2. For each learning period ($\Delta t$ above, e.g., 4 weeks corresponding to a business period of the supermarket) in available POS data, the value of GBE is computed. The method is evaluated in experiments of Sects. 4 and 5, relating the change in GBE to the interpretable visualizations to explain latent changes in consumers' interest, and comparing with baseline methods.

## 2 Explanatory Signs of Changes in Consumers' Behaviors

### 2.1 Four-Step Model of Variety Seeking and Focus Making

A consumer explores various products [15], by such an act as browsing the real space of supermarkets or online shop stores. Then, one may look at the details of an interesting item, until finally deciding to buy it after comparison with other items. By the time of the decision to buy, the consumer may be influenced by peripheral information such as the name of a famous person who used the item [16] or externalizes topics of one's own latent interest via communication with others [17]. All in all, events and information that cannot be included in data may affect consumers' behaviors via providing new contexts of consumption. A context here means a latent condition of any behaviors of the consumers that does not appear in the data.

The outline of the process toward the emergence of contexts is illustrated in Fig. 1. Here, note that Fig. 1 illustrates the outline conceptually, for imaginary (not real) POS data as in Eq. (1). Here, each node represents a purchased item and the closeness of two nodes their tendency to co-occur in baskets. A node colored the more densely shows an item purchased by the higher frequency. Each sub-graph in which nodes are connected via solid lines, shown in an ellipse of a dotted line, is a cluster. A cluster is a set of items that tend to be purchased in the same baskets, as will be defined more specifically in 3.1 for experimental implementation. Referring to Fig. 1, a rough qualitative model of dynamics in the market is summarized as Phase 1 through 4 below. This process borrows its basis from theories of consumers' behaviors including variety seeking [15] and decision making in a dynamic market with uncertain events [18].

(Phase 1)   Consumers are interested in some part of the market, i.e., as in the large cluster of items (i.e., products) in Fig. 1a

(Phase 2)   The interests of various consumers diverge to create a number of clusters, due to their awareness of new contexts, where various items can

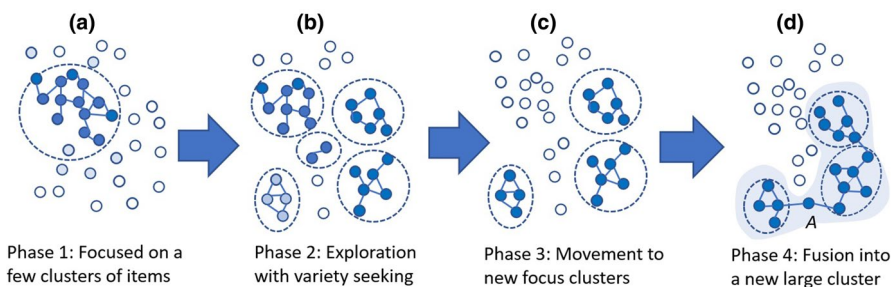

**Fig. 1** Transition of consumers' preference foci. The more densely colored clusters include items by the higher frequency. The separation, movement, and the uniting of clusters correspond to diversity seeking exploration, shifts of interest, and the emergence of new contexts (behind clusters) to fuse consumers' interests

|            | be used or consumed, via their interaction with various information in exploratory activities, as in Fig. 1b |
|---|---|
| (Phase 3) | After exploring clusters, consumers as a group may come to highlight selected clusters as in Fig. 1c, corresponding to a new context in the market that biases consumer's actions to purchase |
| (Phase 4) | The clusters may get united, because each consumer having been interested in some specific clusters of items, changes to buy items from multiple clusters in which other consumers were interested, via communications in the context that emerged in Phase 3. Thus, the united clusters form a new large cluster |

### 2.2 Detecting Explanatory Signs of Changes

Based on the rough model of the consumers' contextual preference transition above, we aim at enabling to explain changes in the market. Here, the explanation of a change in the market means to relate an event, that occurs in the transition from a phase to the next, to previous or forthcoming phases via an admissible (coinciding with other marketers') hypothetical causality. On this explanation, a plan of business can be presented. To explain a change in this sense, it is desired to execute the following steps.

Step 1. Detect an event in which the dynamics of the market can be explained, e.g., a sign (precursor or an observable evidence) of a change in the market.
Step 2. Explain the cause and the effect of the change above.
Step 3. Propose a plan of actions to suppress or enhance the change, expecting any benefit in business, on the explained causality in Step 2.

The event in Step 1, that initiates these three steps, is called an explanatory sign here. I should be noted that explanatory that means a different concept from explainable that is recently studied in machine learning [19]. When one says that X is explainable, it means that X can be explained (usually to humans). On the other hand, when one says that X is explanatory, it means that X is useful information for explaining something, as in the usage "explanatory hypothesis" [20]. In the sense that this event has an influence on Step 3 where a plan of actions is proposed to/by a decision maker(s), an explanatory sign is a type of a "chance" in chance discovery [18]. In addition, for enabling Steps 2 and 3, we visualize a sequence of graphs for periods close to the time an explanatory sign. By visualizing the co-occurrences of items in the data here, the user can relate the latent context (cause) of the consumption represented by each cluster of items (effect). Therefore, the information about an explanatory sign, to be detected in Step 1, should be related to the graphs used in Step 2, so that the sign can be related to latent causes via clusters in the graph. Thus, we developed a method to highlight events appearing when the structure of the graph changes significantly that corresponds to the timing of a change between phrases in the process above. By regarding such an event, as an explanatory sign of change, the user can explain the essential change in the market, e.g., "the interest of consumers has been expanding to various liquors (Phase 2) but is now focusing on a cold beer or on cold wine (Phase 3) because they drink out-doors under the hot sun," or "each customer recently buys various cold drinks (Phase 4)" that leads to the action to sell drinks and foods for reducing the sensible temperature.

The definition above of explanatory signs of changes is beyond a mathematical specification in the form of optimizing an object function computable on given data, because explanatory signs of changes mentioned above are linked to the interpretation of the visualized graphs with relating to causal events not included in data. Therefore, we focus on quantifying the likeliness of an event to be an explanatory sign using the graph-based entropy (GBE) below in the next section, based on the co-occurrence graphs (Step 1). In addition, then, we evaluate the performance of GBE in detecting explanatory signs by comparing with changes in the sales of items in the target category. Then, let us show examples of explanations by real marketers in a supermarket (Steps 2 and 3 above), that will go more to details in the future work.

## 3 Graph-Based Entropy

Kahn suggested using entropy as a measure of variety seeking tendency of consumers [15]. Entropy has also been used in political and marketing sciences to analyze uncertainty and variety in the behaviors of societies and organizations [21–23]. The entropy of each part of an image and its variation has been used for detecting contours and changes in the image [24], exemplified for detecting the precursors of weather change. Furthermore, the entropies of traffics and of events in computer networks turned out to provide a scalable technique to detect unexpected behaviors and abrupt changes [25, 26]. In this paper, graph-based entropy, a quantitative index for evaluating structural changes, is proposed based on clusters obtained as connected sub-graphs in the co-occurrence graph of items in the POS data. Below, let us model the process above, i.e., the transition of consumer behaviors from/to variety seeking to/from focusing interest towards decision making, as the decrease/increase in graph-based entropy. Because a basket of items in POS data as in Eq. (1) is in a similar position to a sentence of words in a document [27], a group of active faults quaking in a consecutive set of earthquakes in a certain period [18], or to a set of stocks whose prices increase in the same week [28], the author expects the presented method can be extended to explaining changes in various application domains.

### 3.1 Graph-Based Entropy as an Index of Explanatory Diversity

Graph-based entropy (GBE) is defined as in Eq. (2).

$$Hg = -\sum_{j} p(\text{cluster}_j) \log p(\text{cluster}_j), \tag{2}$$

where $p(\text{cluster}_j) = \frac{\text{freq}(\text{cluster}_j)}{\sum_j \text{freq}(\text{cluster}_j)}$.

Here, freq(cluster$_j$) denotes the frequency of events in POS data, to which cluster $j$ is the closest among all clusters. An event here means the event that some consumer purchased a set of items in a basket [$B_t$ in Eq. (1)], so a basket corresponds to an event. In defining the closest cluster to a certain basket, the measure of closeness is defined by the cosine of two binary vectors, i.e., $\vec{u}(u_1, u_2, \ldots u_m)$ for basket $u$ and $\vec{v}(v_1, v_2, \ldots v_m)$ for cluster $u$. In $u_i$ and $v_i$, the presence of the $i$th among $m$ items in the market is represented by 1 (the absence by 0). A cluster here means a group of items, connecting the top $\rho N(N + 1)/2$ pairs of the $N$ nodes via edges, corresponding to the $N$ items in the data. The top pairs here mean pairs of the highest co-occurrence, where a co-occurrence is given by the pointwise mutual information, i.e., $p(x$ and $y)/p(x)p(y)$ for a pair of items $x$ and $y$. Here, $p(x)$ is the proportion of baskets including item-set $x$, $\rho$ is the given (as in 3.2) density of the co-occurrence graph to obtain, from which clusters above are taken as connected sub-graphs. In each panel of Fig. 1, a cluster is shown by the set of nodes in a dotted ellipse.

In obtaining clusters, we do not employ projection to the distance space of topics or distributed representation of items, because the aim of this paper to enable explanation defined in Sect. 2 is realized by visualizing item graphs in this paper. In addition, there is a reason why we should take this graph-based clustering rather than the distance-based. If we apply the distance-based method for convex clustering such as $k$-means, the three clusters in Fig. 1 (c) will not be united into one even after they get bridged as in (d) as far as $k$ is held to 3, because the centers of the original three clusters still have stronger gravity than bridges such as node $A$. In addition, there is no obvious logic to reduce $k$. Hence, a tendency of distance-based clustering method to cut such a bridge as node $A$ and disable to organize the large cluster as in (d). Although we may avoid detecting undesired convex clusters by such methods as spectral clustering [29], its computational complexity is O($n^3$).

Because such a bridge may play an important role in forecasting or creating new trends in societies and markets [18, 26], we choose to obtain clusters by connecting nodes via edges representing co-occurrence. Thus, we call Hg in Eq. (2) a graph-based entropy (distinguished from entropy-based clustering [30], where entropy is minimized for clustering). When time $t$ is considered, Hg($t$) is computed by obtaining clusters and frequencies as in Eq. (2) for the data in the time range of [$t − \Delta t, t$]. cps$_{GBE}$($t$) in Eq (3) represents the time derivative of Hg($t$), for periods when the derivative takes a positive value:

$$\text{cps}_{GBE}(t) = \max\left\{\text{average}_{0<dt\leq\Delta t}\text{Hg}(t - dt) - Hg(t), 0\right\}. \quad (3)$$

Here, cps$_{GBE}$($t$) means that the decrease in Hg($t$) from the average value in the recent period of length $\Delta t$. cps$_{GBE}$($t$) takes the larger value if consumers' interest had ranged across various clusters of items in the market at time $t − dt$, followed by focusing on restricted clusters of items at time $t$. Thus, a larger positive value of cps$_{GBE}$($t$) means the combination or focusing of consumers' interest in fewer clusters. Here, in this paper, the value of cps$_{GBE}$($t$) is regarded as the change-point score obtained by GBE, for time $t$. That is, cps$_{GBE}$($t$) is an index of the structural change of consumers' interest is large, from the exploration of various items to the choice of focused items, that corresponds to Phases (c) and (d) in Fig. 1. Let us hereafter focus on these two phases that are more noteworthy in detecting decision making of consumers than Phase (a) or (b).

The direct reflection of changes in the graph structure is a feature of GBE that differentiates it from the previous methods for change detection. Among them, Local Linear Regression (LLR [7]) is taken as a baseline in the experimental comparison in Sects. 4 and 5. On the other hand, in comparison with the dynamic topic models (e.g., DTM [8]) to be also compared experimentally, GBE is free from the constraint of DTM that each of the $K$ topics should succeed one of the topics in the previous periods, because clusters in GBE can be separated/united to lose/form topics fitting the contexts in the external world.

These differences of GBE from existing approaches mentioned above are illustrated in Fig. 2. The transition of Hg($t$), in the right of Fig. 2, differs from that of topic distributions in dynamic topic models in that the contexts (corresponding to topics in topic models) in GBE are born or lost via their separations or uniting of graph-based clusters, adapting sensitively to the dynamic interests of consumers due to the emergence of new context formed via such a bridge as node $A$ in Fig. 1. Experimental comparison with change-point scores obtained by other methods will be shown experimentally in Sects. 4 and 5.

### 3.2 The Algorithm to Compute GBE and Its Time Derivative

Based on the definition of GBE and cps$_{GBE}$ above, Algorithm 1 for detecting a change in the structure of the co-occurrence graph is shown below. The main function calls function Hg($t$) corresponding to the computation in Eq. (2). The top $\theta_r$ times, i.e., the times of the $\theta_r$ largest cps$_{GBE}$($t$) in Eq. (3), are taken as times when signs of structural changes are detected.

The data given is represented by $D$, where $T$ is the length of the time series, in which each time $t$ includes $B_t$ that is a set of baskets at $t$ that is really a period such as a week. A basket is a set of items, and each item starts from being one cluster including only itself, in initializing clusters. Clusters are obtained by connecting items co-occurring frequently with each other, where the co-occurrence of two items item$_a$ and item$_b$ in set $B$ of baskets is computed as the pointwise mutual information [31] as in Eq. (4):

$$\text{cooc}_B(\text{item}_a, \text{item}_b) =: \frac{p_B(\text{item}_a, \text{item}_b)}{p_B(\text{item}_a) p_B(\text{item}_b)}. \tag{4}$$

In Eq. (4), $p_B(\text{item}_a)$ represents the probability of the occurrence of item$_a$, computed as the number of baskets including item$_a$ divided by the number of baskets in $B$. In Algorithm 1, cooc$_{B(t-\Delta t):B(t)} \cdot \text{rank}(\text{item}_i, \text{item}_j)$ in line 11 means the rank of pair $\{\text{item}_i, \text{item}_j\}$, ranked on the co-occurrence in the basket sets from time $t - \Delta t$ through time $t$. Here, $\Delta t$ should be set to a period of the periodical behavior of the market, e.g., a month that is a unit of business period for the supermarket. The highest $\rho|I|^2/2$ pairs of items, in the ranking of co-occurrence, are connected by edges

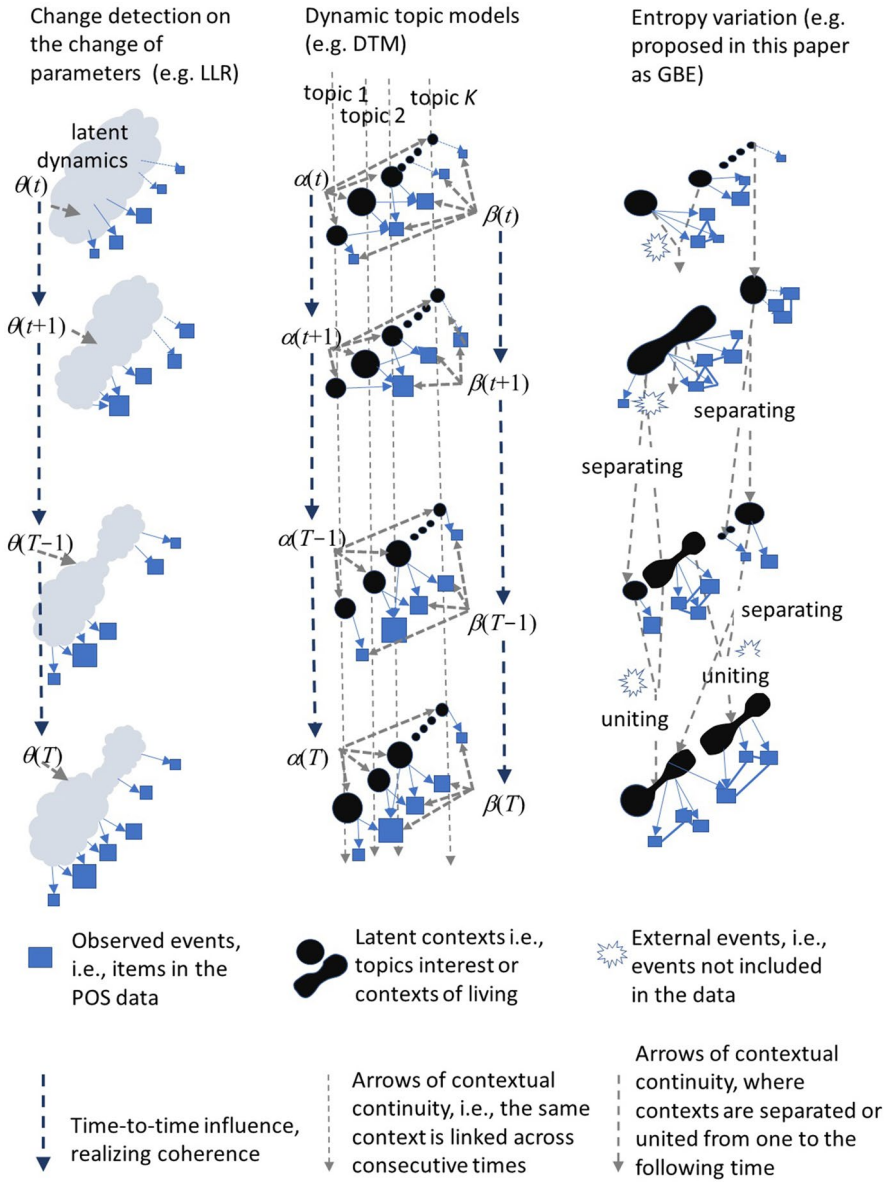

**Fig. 2** Three methods compared: graph-based entropy in the right of this figure differs from the other methods in the sense that the appearance, the separation, the disappearance, and the uniting of contexts, underlying clusters, are reflected on the reformation of graph-based clusters (illustrated by black spots) of items

to form a cluster, so that the density of the graph, represented by $\rho$ as stated in 3.1, is set constant. The value of $\rho$ is set to 0.06 here, by which the number of clusters came to be between 10 and 20 for all the given categories, that makes the graph easy to overview and interpret for marketers when visualized as in Sect. 5.

Two items of high co-occurrence take the same clusterID and combined to form one cluster, i.e., a connected sub-graph, as in line 12. Each cluster is reflected on Hg(*t*) in line 15. Then, cps$_{GBE}$(*t*), the time derivative of Hg(*t*) in Eq. (3) is obtained

**Algorithm 1: Detecting structural changes with GBE**

**In:** $D = \{B_t | 0 < t \leq T\}, I$ = all items in *D*
**Out:** change point score cps$_{GBE}$, the sign of change alert$_{GBE}$(*t*)
1: **initialize clusters:**
2:   $\forall$ item$_i \in I$, clusterID(item$_i$) = *i*
   $\Delta t$ = constant.
3: **for** *t* in [0, *T*] **do**
4:   cps$_{GBE}$ = *d*Hg(*t*) *for* $\Delta t$ in Eq. (3)
5:   **if** rank$_{t \in [0,T]}$ cps$_{GBE}$(*t*) < $\theta_r$   alert$_{GBE}$(*t*) = Truth
6:   **else**         alert$_{GBE}$(*t*) = False
7:   **end if**
8: **end for**
9: **function: compute the graph-based entropy** Hg(*t*):
10: **for** item$_i$, item$_j \in I | i \neq j$,
11:   **if** cooc$_{B(t-\Delta t):B(t)}$·rank(item$_i$, item$_j$) $\leq \rho |I|^2/2$ **do**
12:    clusterID(item$_j$) = clusterID(item$_i$)
13:   **end if**
14: **end for**
15: **return** Hg(*t*) in Eq.(2) for clusters of non-empty clusterIDs

## 4 Experiments

Here, let us show results of comparison executed between (1: rank change) that is weekly change in the ranking of items in their purchase frequency (2: LLR) the Local Linear Regression for the detection of changes, (3: DTM), and (4: the time derivative of GBE) cps$_{GBE}$(*t*), for each category of items.

Let *D* be the data on purchases of items (item classes precisely, as mentioned just later) in a given target category of items in the market, including a sequence of $B_t$ meaning a purchased set of baskets each time *t*, that really means 4 weeks ending with the *t*th week in the data, as in Eq. (5), where $L_t$ is the number of included baskets. For example, let us define *liquor* as the target category. Then, items are really item classes, for example, wine of 720 ml bottle, beer of 350 ml, etc. Here, the author does not deal with each product item given as Casillero del Diablo 720 ml bottle, Kirin draft beer of 350 ml, etc:

$$B_t := \{\text{basket}_1, \text{basket}_2, \ldots, \text{basket}_{L_t}\}. \tag{5}$$

POS data for 1 year for four retail stores have been provided by Kasumi Co. Ltd, a supermarket chain. Here, I deal with ten categories of items that are dry food, stationary, liquor, ice cream, bread, spice, sauce, vegetable, processed food, and processed meat. The POS data from each store dealt with here had 35,995 baskets (seven the ten categories above), 49,853 baskets (8 categories), 22,299 baskets (nine categories), and 17,607 baskets (six categories), on the completeness of the data. That is, stores lacking in items of some subcategories of a category are excluded from the target data, to assure the unique assignment of subcategories to the same items. We divided all baskets in the year in each store into weeks, so $T$ was 52 weeks.

In computing Hg($t$) and cps$_{GBE}$($t$) in Eqs. (2), (3), $\Delta t$ is set to 4 weeks, for aiding marketers in a supermarket to explain the signs of changes, because 1 month is a unit of business time and weeks are regarded as units of consumers' life. As mentioned in Sect. 5, $\Delta t$ is partially set to 2 weeks just for experiment.

**(Baseline 0: Regarded as Correct Changes) Weekly Changes in the Rank of the Frequency of Items.** The rank change computed here is regarded as the real change, because this directly reflects consumers' preference change. For $B_t$, the $R$ top items of sales volume, i.e., the number of baskets including each item, are sorted as in Eq. (6). Here, ranked$_i$($t$) is the ($i+1$)th most frequent item in basked set $B_t$, i.e., at time $t$:

$$\text{top}_R(t) := \{\text{ranked}_0(t), \text{ranked}_1(t), \ldots, \ldots \text{ranked}_{R-1}(t)\}. \quad (6)$$

Here, ranked$_i$($t$) represents the $i$th most frequently purchased item at time (week) $t$, and top$_R$($t$) the set of ranked$_i$($t$) for all $i$ in [1, $R$]. Then, $t$ is regarded as a time of rank change that means a substantial weekly change or a precursor of change, iff alert$_{real}$($t$) is Truth as in line 2. That is, a rank change is detected at time $t$, iff the total weekly change of ranks of items, i.e., cps$_{real}$ ($t$) obtained in line 13, becomes larger than its average of 3 weeks before $t$ as in line 1. In addition, as in line 1, an event at or before a rank change by within 2 weeks is regarded as a candidate of the precursor.

**Algorithm 2: detecting the rank change of items**
**In:** $D = \{B_t | 0 < t \leq T\}, I$ = all items in $D$
**Out:** cps$_{real}$ ($t$) and alert$_{real}$($t$)
1: **if** $\exists t' \in [t, t+2]$, cps$_{real}$ ($t'$) > avr$_{t \in [t-3, t-1]}$ cps$_{real}$ ($t$)
2:    **then** alert$_{real}$($t$) = Truth
3:    **else** alert$_{real}$($t$) = False
4: **end if**
5: **Compute the total rank change cps$_{real}$($t$)**
6:    **for each** ranked$_{i \in [0, R-1]}$($t$) $\in I$ **do**
7:       **if** ranked$_i$($t$) $\in$ top$_R$($t - 1$) **then**
8:          $j = \min_{\text{ranked}_i(t) = \text{ranked}_{i'}(t)} i'$
9:          change$_i$($t$) = ($R - i$)|$i - j$|
10:    **else** change$_i$($t$) = $R(R - i)$
11:       **end if**
12:   **end for**
13: cps$_{real}$ ($t$) = $\sum_{i \in [0, R-1]}$ change$_i$($t$)

Intuitively, the weekly change at $t$ represented by $\text{cps}_{\text{real}}(t)$ in line 13 means the extent to which $\text{top}_R(t)$ differs from $\text{top}_R(t\text{-}1)$, putting weights on items of higher rank at time $t$ (by the factor $R-i$ to the $i$th highest). For example, if $\text{top}_3(t\text{-}1)$ was {apple, banana, orange} and $\text{top}_3(t)$ is {tomato, apple, orange}, the distance of movement of apple from time $t$-1 to time $t$ is counted to be 1 that is the absolute value of 1 ($=i$) minus 0 ($=j$) in the RHS of line 9. On the other hand, the movement distance of tomato as $\text{ranked}_0(t)$ (the first in time $t$) is 3 that is factor $R$ in line 10, because it did not appear in $\text{top}_3(t\text{-}1)$. As a result, the appearance of a new item causes a greater influence on $\text{cps}_{\text{real}}(t)$ than the movement of items that existed since the previous time. Furthermore, the factor $R-i$ in the RHS of lines 9 and 10 is used to highlight the movement of an item of the higher rank at time $t$.

We take the total rank change $\text{cps}_{\text{real}}(t)$ in line 13 as the real change of week $t$, reflecting the requirements of marketers of supermarket chain Kasumi Co. Ltd. that noteworthy changes are the shifts of ranks, especially of higher ranked items.

**(Baseline 1) Local Linear Regression.** Whereas the previous studies on change detection often assumed that changes occur abruptly [5, 6], changes came to be assumed to really take place continuously in Local Linear Regression (LLR [7]). In the market, consumers may seem to change suddenly if affected by external events or information such as TV programs, news, or opinions in SNS. However, the information really makes an influence, continuously taking time, on the process of consumers' decision making rather than to their actions directly and suddenly. By detecting times of high values of the original change-point score, specifically defined for LLR, changes are detected with high accuracy at an early moment after the starting of the real change, i.e., within a short tolerant delay. LLR has been experimentally shown to be effective for real-life data on the events in servers, industrial machines, etc. In this paper, the same 1-year data, as given to other methods, has been given to LLR, in the form below for all the $T$ weeks. In Eq. (7), $p_{\text{item}i}(t)$ represents the proportion (in the range of [0, 1]) of baskets in which item $i$ was bought in the $t$th week:

$$\text{input}_{\text{LLR}}(t) := \left(p_{\text{item1}}(t), p_{\text{item2}}(t), \ldots p_{\text{item}\in I}(t), \ldots p_{|I|}(t)\right). \quad (7)$$

In LLR, the change-point score is computed, setting extinction coefficient $r$. For example, if $r$ is 0.7, i.e., the influence of weak $t$-1 is reflected on the learning of the parameter $\Theta(t)$ weakened by the factor of $1-r$ as 0.3, so that we regard 4 weeks ago as ignorable due to the factor of 0.01. Thus, this condition enables a fair comparison with GBE, if $\Delta t$ for GBE [the time window of data for computing $\text{Hg}(t)$] is set to 4 weeks. Thus, $r$ has been set to 0.7 in the $t$ test in Sect. 5. To compare with other methods, let us refer to the change-point score of LLR at time $t$ as $\text{cps}_{\text{LLR}}(t)$.

**(Baseline 2) Dynamic Topic Model (DTM).** DTM introduced in the introduction has been here applied to the POS data assuming that the purchase of each item is caused by consumers' interest in a given number of topics. By DTM, we can model the transition of topic distribution of each item as time passes by reflecting the time-to-time continuity. In this experiment, DTM has been employed as a tool for detecting changing points of consumers' behaviors with a dynamic model of latent structural causality. The number of topics has been set to the range of [3, 10] and parameter $\alpha$ to 0.1, where the number of peaks for the tested 1 year came to be comparable to compared methods. Here, Eq. (8) has been taken as the change-point score:

$$\text{cps}_{\text{DTM}}(t) = 1 - \cos(\theta(t), \theta(t-1)), \quad (8)$$

where $\theta(t)$ is the topic vector of the *t*th week, counting each unique item in one basket purchased as each unique word in one document in DTM [8]. The topic-tracking model (TTM [9]) has also been compared, but we choose only DTM to show experimentally in this paper, because our POS data did not include customers ID for all purchases that is a part of input data for TTM. In general, a supermarket tends to have a lot of customers having no customer IDs. DTM has not such condition and can be tested free from the package provided in https://github.com/magsilva/dtm/tree/master/bin.

For the 1-year sequence of events, the data of 4 weeks (as $\Delta t$) are used for evaluating $\text{cps}_{\text{GBE}}$ for each week. These data are used for obtaining $\text{cps}_{\text{GBE}}(t)$ on the time derivative as far as Hg(*t–dt*) could be obtained for *dt* in range of [1, 4] as in Algorithm 1. We can point out the convenience of GBE in that it runs for smaller data (as well as large data) than LLR or DTM using full data for evaluating the change-point scores of each week.

## 5 Results and Discussion

Based on the experimental results, let us here present the features of GBE from three aspects: in 5.1, we choose two stores of the supermarket chain and the visible correspondence of the curves of change-point scores to the real rank change. In 5.2, the curves are related to co-occurrence graphs of items, for explaining causalities including events out of POS data. In 5.3, statistic comparison with baselines for all the four stores.

### 5.1 The Correspondence of Change-Point Scores with the Real Rank Change

The three change-point scores, i.e., $\text{cps}_{\text{GBE}}(t)$, the $\text{cps}_{\text{LLR}}(t)$, and $\text{cps}_{\text{DTM}}(t)$, were compared here with the real rank change $\text{cps}_{\text{real}}(t)$. In Fig. 3, the curve for the category of "cooking spice" in store 1 is shown as an example. In (a), the comparison is made for all weeks in the target year. In (b), the period from the 7th until 12th weeks is extracted from (a).

In Fig. 3, the three methods $\text{cps}_{\text{GBE}}$, $\text{cps}_{\text{DTM}}$, and $\text{cps}_{\text{LLR}}$ have peaks from the 28th until 31st weeks in category "cooking spices" of store 1, where $\text{cps}_{\text{real}}$ increases with some turbulence. The peak of $\text{cps}_{\text{real}}$ in the 10th week fits the peak of $\text{cps}_{\text{GBE}}$. $\text{cps}_{\text{GBE}}$, $\text{cps}_{\text{DTM}}$, and $\text{cps}_{\text{LLR}}$ is rising from the 46th week that is the peak of the real change and find peaks in the 47th week. Thus, the peaks of the three functions tend to synchronize with the curve of $\text{cps}_{\text{real}}$ that is here regarded as the real change of consumers' preference. In Fig. 4, the curve for the category of "bread" in store 2 is shown. The peaks of $cps_{GBE}$, $cps_{GBE}$, and $cps_{DTM}$ do not coincide in this case as well as in (a), so let us expand a part of the 30th through the 46th week as in (b). Here, the peak of $cps_{real}$ in the 39th week is preceded by the peak of $cps_{GBE}$ in the 37th week. In addition, the increase in $cps_{real}$ after the 42nd week is preceded by the peak of $cps_{GBE}$. Neither of these peaks coincides or is preceded by the peaks of $cps_{DTM}$ or $cps_{LLR}$.

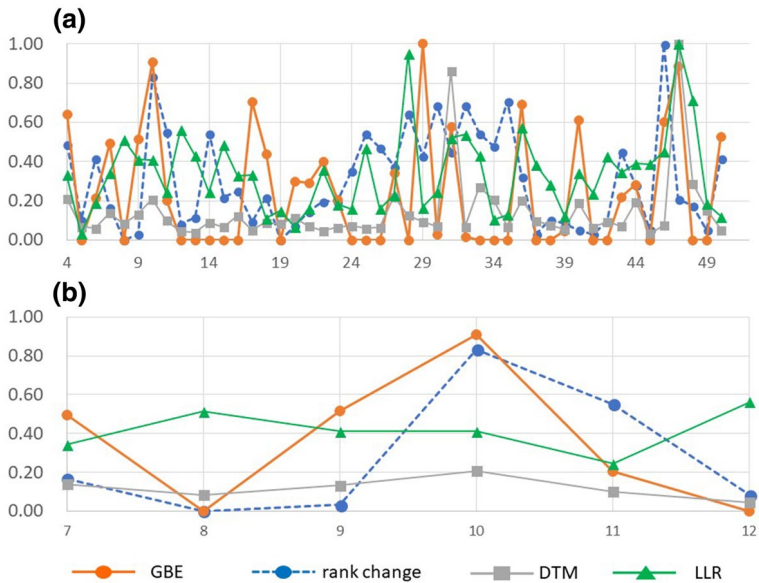

**Fig. 3** Real rank change (dotted line) in the category of "cooking spice" for store no. 1 and the change-point scores on GBE, DTM, and LLR (solid lines: $cps_{GBE}$, $cps_{DTM}$, $cps_{LLR}$), for the 4th–50th weeks (**a**), and the extracted 7th to the 12th week (**b**)

In summary, some visible correspondence of the change-point scores with the real rank change is found and $cps_{GBE}$ shows relative strength in detecting changes and their precursors. Although obvious superiorities of $cps_{GBE}$ are not always found by just looking at the curves, $cps_{GBE}$ precedes $cps_{real}$ in some cases. Let us evaluate the feature of GBE integrating a user-oriented viewpoint below.

### 5.2 Relating the Curve and Co-occurrence Graphs of Items for Explaining Causalities, Considering External Events

Here, let us exemplify the proposed general method in Sect. 2.2, to relate the time series curve of change-point score and the co-occurrence graphs, so that a marketer can explain the latent causality of changes. That is, the user who is supposed to be a marketer first detects the periods of high change-point score, for which the relations of events are selected and visualized. As a result, a sign of a change in the market may be detected (Step 1). Then, he/she explains the causes and the effects of the change (Step 2), to propose a plan of actions to suppress or enhance the change, expecting benefits in business (Step 3).

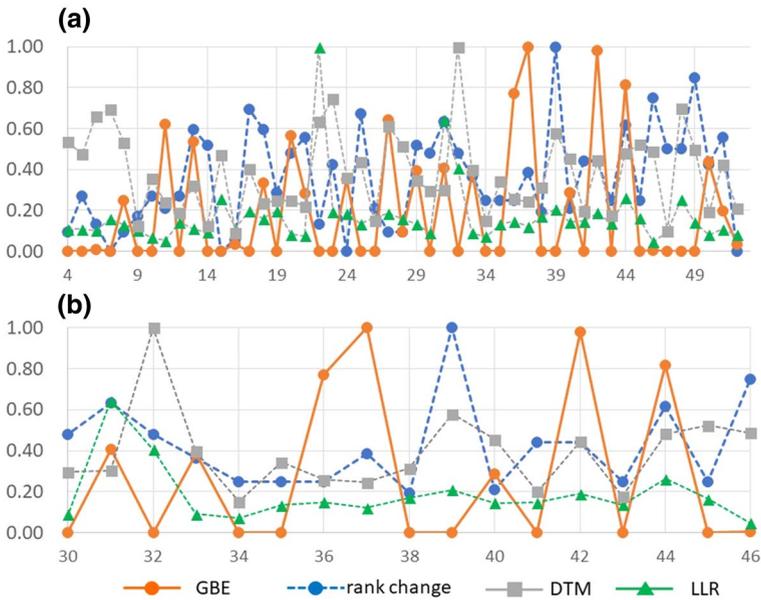

**Fig. 4** Real rank change in the category of "bread" of store no. 2 and the change-point scores on GBE, DTM, and LLR (solid lines: cps$_{GBE}$, cps$_{DTM}$, cps$_{LLR}$), for the 4th–50th weeks (**a**), and the extracted 30th–46th weeks (**b**)

Corresponding to (b) in Fig. 3, the transition of the co-occurrence graph of items is shown in Fig. 5 for the same data. Here, the nodes represent 20 item classes in the category. In addition, $\rho|I|^2/2$ pairs (in Algorithm 1), pairs of nodes, are connected via edges. The dotted lines show singly connected lines and the solid multi-connected. That is, if any edge of the dotted line is cut, the graph gets separated into connected sub-graphs, i.e., clusters. A sub-graph connected by either solid or dotted lines forms a cluster.

In the sequence of graphs in Fig. 5 for category "cooking spices", we find the structure of the graph starts to change from the 9th week, by changing the links of the cluster in the lower part and involving "cream stew" into it. In addition, the cluster gets separated into two clusters in the 10th week. Then, the cluster in the lower half of the graph is reinforced in the 11th week, as shown by the generated cluster. These times of structural changes, i.e., the 9th and 10th weeks, coincide more obviously with peaks of cps$_{GBE}(t)$, as shown in Fig. 3, than with cps$_{LLR}(t)$ or cps$_{DTM}(t)$. "Cream stew" is found to stay in the finally reinforced (multi-connected) cluster in graphs from the 8th until 11th weeks, and other spices in this cluster are also used in cooking stew. Using Google Trends (https://trends.google.co.jp/trends/) for the Japanese query "shichu" that means "stew", as in Fig. 6, we find that the interest of people in eating stew gets highlighted from the latter half of August every year that nearly coincides with the 9th week when "cream stew" joined the lower cluster. On this finding, the marketers of the supermarket found new actionable plan to promote foods and drinks relevant to stew that are not only spices but also side foods (e.g., bread, pickles, and cabbage), and food to put in stew (potato, onion, mushrooms, and meat), and also such tools as stewpans for cooking stew. This result means that if the marketer first uses a curve in Fig. 3 to choose the period in Fig. 5 (Step 1), to explain the causality of the change (Step 2), and to propose actions of business (Step 3), $cps_{GBE}(t)$ works in aiding his/her process better than the compared methods.

In addition, in the case of "bread" corresponding to Fig. 4, for the period from the 34th until the 44th weeks, corresponding to the changes in the last part of Fig. 4 (b), the transition of the graph is shown in Fig. 7. Here, the nodes represent the 20 items in the category. In the sequence of graphs here, we find that the structure of the graph starts to change substantially from the 37th week and the cluster of "Chinese" bread (mantou with meat inside that is classified in the category of bread in this supermarket) is suppressed. Then, from the 42nd week, the new small cluster including "croissant" appeared and stayed in the graph. These structural changes in the 37th and 42nd weeks coincide with the peaks of only $cps_{GBE}(t)$, among the three in Fig. 6. The finally created cluster shows breads and bans used in parties with friends and families that are popular in this season (the end of March) in the Japanese culture, because cherries blossom and attract people

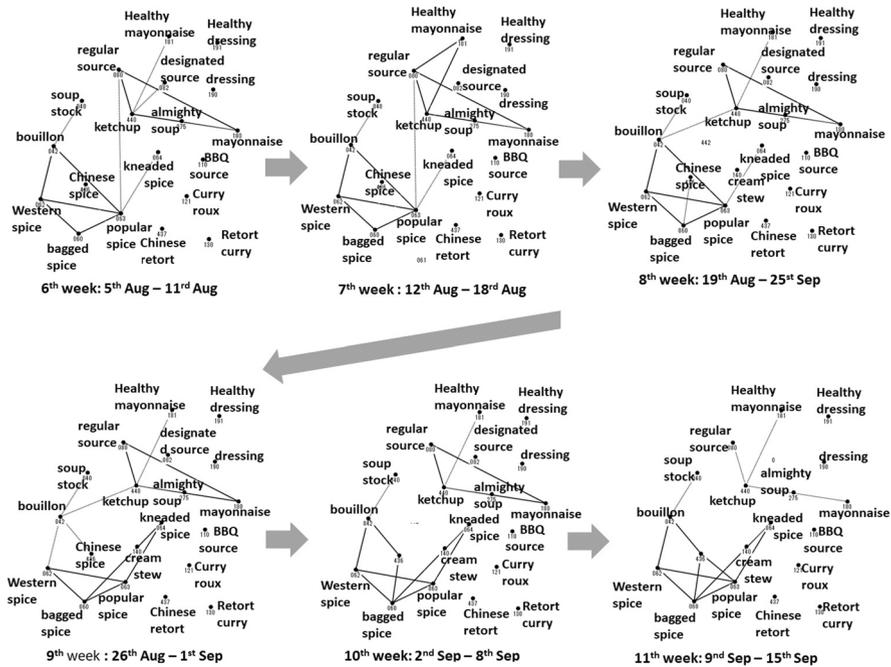

**Fig. 5** Variation of the graph corresponding to the 6th through the 11th week for rathe category "cooking spices" in store 1. In the 9th and 10th weeks corresponding to the changes of GBE in Fig. 3, the structure of the graph changes

to do parties under cherry trees, and the new year of schools and firms starts from April. Such a culture had been known to the marketers of the supermarket who provided the POS data, but the importance of the new cluster has not been recognized so far. The combination of detecting changes with GBE and the visualization of graphs thus aids marketers' insights. The marketers of the supermarket found a new actionable plan to promote sales of bread and other foods and drinks (e.g., cakes and wines) with advertisements to relate those items with parties of young people.

### 5.3 Statistic Comparison with Baselines

The performance of detecting the signs of change, i.e., of changes or of their precursors, has been evaluated by a statistic comparison. Here, the correspondence of the times of the top values of $cps_{GBE}$, $cps_{LLR}$, $cps_{DTM}$, and $cps_{real}$ has been evaluated. The timing of the rank change is given by the times when $alert_{real}$ is True in Algorithm 2. The top values of $cps_{GBE}$ are defined by such times when $alert_{GBE}$ is Truth in Algorithm 1. In addition, the times of the same number ($\theta_r$ in Algorithm 1) of the highest values for $cps_{LLR}$ and $cps_{DTM}$. In the evaluation below, the precision of method $M$ is computed as the proportion of $t$, where $alert_{real}$ is True, among all $t$ of the top values of $cps_M$. The recall of method $M$ is the proportion of times of the top values of $cps_M$, among all $t$, where $alert_{real}$ is True.

As a result, the precision of GBE is significantly larger than LLR and DTM as in Tables 1 and 2a. In Table 2, $r$ of LLR has been set to 0.7 in the $t$ test, as mentioned in Sect. 2.2. LLR and DTM have no significant difference, implying that the previous methods did not show such a significant improvement in the detection of precursors as GBE shows. The recall of GBE is also higher than LLR and DTM as in Table 1, although the superiority against DTM is not as significant as of precision according to Table 2b.

An observed phenomenon of DTM in this experiment was that the evaluated recall, precision, and F1 are not monotonically larger for the larger number of topics. In more detailed observation of the analysis by DTM, beyond the results in Tables 1 and 2, items in small (low probability) topics sometimes switch with other small

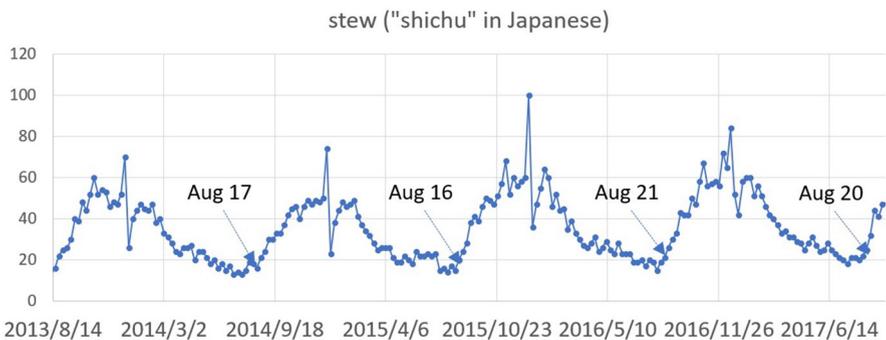

**Fig. 6** Changes in the popularity of stew according to Google Trends search

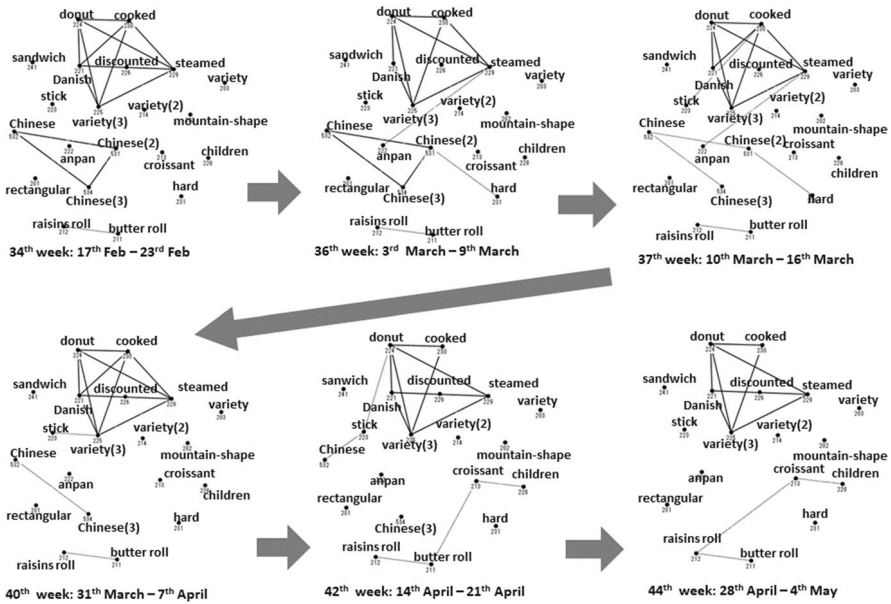

**Fig. 7** Graphs for the 34th through the 44th weeks for the category "breads" in store 2. In the 37th corresponding to a peak of cps$_{GBE}$(*t*) in Fig. 4, the structure changed

topics, for 1–3 weeks around the top values of change-point score. As a result, the changing moment was not obtained stably. Such a tendency may not be found in evaluating DTM on such standard criteria as perplexity because switching of all words in a topic with another topic does not affect perplexity substantially. In summary of the results, we can say GBE is showing a breakthrough in finding high precision signs of changes, which are sometimes precursors of changes in consumers' purchase priority.

## 6 Conclusions

GBE is presented here, based on a model of consumer's preference shift, that go via preference diversity to focusing. The experiments first show the high correspondence of the change-point scores, obtained on GBE, to the correct real changes in the market. Furthermore, the result supports the proposed method to detect the peaks of change-point score on GBE and take the co-occurrence graphs corresponding to the times of those peaks, to aid a marketer in explaining the latent dynamics and causalities of changes in the market.

The simplicity of the computing algorithm and its linkage to the structure of items' co-occurrence graph enables a user, who is supposed to be marketers in this paper, to explain the dynamic changes in the contexts behind data on consumers' buying for living. Thus, we can validate the importance of each change both

**Table 1** Comparison of precision, recall, and F1 for GBE and the baselines (LLR and DTM) for the same data of 30 categories for the 4 supermarket stores

|  | GBE $\Delta t=4$ | GBE $\Delta t=2$ | DTM $k=3$ | DTM $k=5$ | DTM $k=10$ | LLR $r=0.2$ | LLR $r=0.5$ | LLR $r=0.7$ |
|---|---|---|---|---|---|---|---|---|
| Precision |  |  |  |  |  |  |  |  |
| Average | **0.521** | 0.512 | 0.486 | 0.485 | 0.461 | 0.430 | 0.466 | 0.453 |
| Std.dev | 0.082 | 0.105 | 0.101 | 0.090 | 0.110 | 0.112 | 0.103 | 0.099 |
| Recall |  |  |  |  |  |  |  |  |
| Average | 0.500 | **0.503** | 0.482 | 0.496 | 0.465 | 0.430 | 0.456 | 0.476 |
| Std.dev | 0.083 | 0.103 | 0.065 | 0.093 | 0.121 | 0.116 | 0.101 | 0.094 |
| F1 |  |  |  |  |  |  |  |  |
| Average | **0.506** | 0.499 | 0.483 | 0.488 | 0.459 | 0.429 | 0.459 | 0.460 |
| Std.dev | 0.077 | 0.079 | 0.097 | 0.086 | 0.110 | 0.110 | 0.097 | 0.088 |

The underlined and bold values are the best performance in each measure

Here, parameters denote, respectively, $\Delta t$ the number of weeks for computing $cps_{GBE}$, $k$ the number of topics in DTM, and $r$ extinction coefficient $r$ in *LLR*

**Table 2** Comparison in the measure of precision and recall of GBE, LLR, and DTM for parameters of the best performance

| X \ Y | (a) Precision | | (b) Recall | |
|---|---|---|---|---|
|  | GBE$\Delta t$=4 | LLR $r=0.7$ | GBE$\Delta t$=4 | LLR $r=0.7$ |
| LLR: $r=0.7$ | **0.014** |  | **0.04** |  |
| DTM: n=5 | **0.059** | 0.23 | 0.27 | 0.11 |

The cells show the *p* values for one-sided *t* test

Each *p* value shows the significance of the superiority of X to Y, for each cell quantitatively (evaluation of the change-point score), qualitatively (explaining the meaning of changes on the graphs), and quickly (detecting signs some of are found to be precursors). The method is currently being introduced to the supermarket having provided the data to the author, for understanding consumers and improving marketing strategies.

**Acknowledgements** This study was funded by JST-CREST Grant Number JPMJCR1304, and JSPS KAKENHI Grant Numbers JP16H01836 and JP16K12428, Kasumi Co. Ltd., and Kozo Keikaku Engineering Inc.